\documentclass[letterpaper,preprintnumbers,prd,nofootinbib,nobibnotes,showpacs]{revtex4}
\usepackage{graphics,epsfig,subfigure}
\usepackage{color}
\usepackage{amsfonts}
\usepackage{mathrsfs}
\usepackage{epsfig}
\usepackage{graphicx}%
\usepackage{dcolumn}
\usepackage{amsmath}
\usepackage{color}
\usepackage{color}

\newcommand{\blue}[1]{\textcolor{blue}{{#1}}}
\begin{document}
\renewcommand{\baselinestretch}{1.3}
\newcommand\beq{\begin{equation}}
\newcommand\eeq{\end{equation}}
\newcommand\beqn{\begin{eqnarray}}
\newcommand\eeqn{\end{eqnarray}}
\newcommand\nn{\nonumber}
\newcommand\fc{\frac}
\newcommand\lt{\left}
\newcommand\rt{\right}
\newcommand\pt{\partial}

\allowdisplaybreaks

\title{A small cosmological constant from a large number of extra dimensions}

\author{Changjun Gao\footnote{Email: gaocj@bao.ac.cn}}
\affiliation{ 1.\quad National Astronomical Observatories, Chinese Academy of Sciences, 20A Datun Road, Beijing 100101, China; gaocj@nao.cas.cn\\
\quad 2. School of Astronomy and Space Sciences, University of Chinese Academy of Sciences, 19A Yuquan Road, Beijing 100049, China}

\begin{abstract}
\baselineskip=0.6 cm
\begin{center}
{\bf{ABSTRACT}}
\end{center}

In this article, we consider the $4+n$ dimensional spacetimes among which one is the four dimensional physical Universe and the other is an n-dimensional sphere with constant radius in the framework of Lanczos-Lovelock gravity. We find that the curvature of extra dimensional sphere contributes a huge but negative energy density provided that its radius is sufficiently small, such as the scale of Planck length. Therefore, the huge positive vacuum energy, i.e. the large positive cosmological constant is exactly cancelled out by the curvature of extra sphere. In the mean time the higher order of Lanczos-Lovelock term contributes an observations-allowed small cosmological constant if the number of extra dimensions is sufficiently large, such as $n\approx{69}$.

\end{abstract}

 \maketitle
\newpage
\section{Introduction}
The first ideas of existence for new spatial dimensions beyond our four dimensional sapcetime can be dated back to the early works of Kaluza and Klein \cite{KK:1921}. They tried to unify Maxwell theory with Einstein gravity by proposing a model with a compact fifth dimension where the Maxwell theory was originated.  However, this theory
 is mired with problems, some of which are explored by Wesson in \cite{wess:1998}. On the other hand, as one of the most important and fascinating theory in modern theoretical physics, string theory \cite{witten:1995} suggests that our universe must be higher dimensional and the compact extra dimensions may be as small as the Planck length and therefore nearly impossible to detect. To this day,  the existence of extra dimensions are necessary for the consistency of the modern physical theories \cite{green:1987}. 
 
 Lanczos-Lovelock (Thereafter, we use $\mathrm{LL}$ to represent it) gravity \cite{lan:1932,love:1971} represents a natural and elegant generalization  of Einstein’s theory of gravity from four dimensions to higher dimensions. It is characterized by the fact that the field equations always contain up to second derivatives of the metric even though the action functional can be a quadratic or very higher degree polynomial in the Riemann curvature tensor. This is very different from the Einstein gravity where only a linear term of Riemann curvature is in the presence. It is found the resulting LL tensor (generalized Einstein tensor) contains the Einstein tensor as a particular case, and it reduces to the latter when the spacetime goes back to four dimensions. The subtlety of second-order differential equations of motion lies in that the linearized LL gravity (around Minkowski space) is ghost-free \cite{zum:1986,zwi:1985}. Furthermore, the number of gravitational degrees of freedom in LL  gravity turns out to be exactly that in Einstein gravity \cite{buch:1991}. In higher dimensions than four, the LL gravity differs from Einstein gravity and brings us an even richer physical landscapes \cite{pad:2013,chris:2009,simon:1988,par:2006}. 

 For cosmology, Deruelle and Fari$\Tilde{\mathrm{n}}$a-Busto constructed systematic cosmological solutions in the LL gravity. These solutions cover the maximally symmetric space-times, the Robertson-Walker universes and the product manifolds of symmetric subspaces \cite{der:1990}.  Mena Marugan \cite{mena:1992} considered a class of higher dimensional LL gravity in the presence of a positive cosmological constant whose metric is given by the product of the four-dimensional physical universe and an n-dimensional  maximally symmetric space. It was shown that when all the LL coefficients
 are non-negative,  the theory admits classical solutions with a constant scale
extra space, and a four-dimensional Einstein gravity with a non-negative cosmological constant. Kirnos and Makarenko \cite{kirnos:2010} studied the accelerating cosmologies in LL gravity with dilaton field. The metric is also assumed to be the product of a four-dimensional universe plus an n-dimensional maximally symmetric extra space. It is found that there are stationary, power-law, exponential and exponent-of-exponent forms of cosmological solutions.  Paul and Mukherjee \cite{paul:1990} investigated a 10-dimensional
cosmology with the Gauss-Bonnet term.  By this way, they proposed a scenario which includes a spontaneous compactification of extra space and an inflationary
four-dimensional universe followed by a radiation-dominated epoch. Furthermore, the cosmological constant problem is argued to be solved by the dynamics of compact space.  Mena Marugan \cite{mena:1992b} developed the perturbative formalism for the LL gravity in higher dimensional spacetime with the induced metric given by the
 product of two maximally symmetric spaces.  A good feature of this formalism is that it admits the study of multidimensional
 cosmological models of physical interest, in which all except four dimensions could become compactified. Pavluchenko \cite{pav:2015}  performed a stability analysis for exponential solutions in LL gravity with  the Gauss-Bonnet term and the cubic term, respectively. It is found that the isotropic universe
 is always stable provided that the Hubble parameter is positive. Pavluchenko \cite{pav:2009} derived the equations of motion for Bianchi-I cosmological models in the LL gravity. The equations derived apply to any number of spatial  dimensions and any order of the LL corrections. M$\ddot{\mathrm{u}}$ller-Hoissen \cite{muller:1986} 
 studied the LL gravity up to the third order in the Riemann curvature.  The equations of motion are elaborated in the Kaluza-Klein cosmology ansatz and solved in the special case where the extra dimensions appear as an n-dimensional sphere with constant radius. Rubakov and Shaposhniko \cite{ruba:1983} discussed the possibility that the cosmological constant problem is solved by raising the number of spatial dimensions 
in Einstein gravity but with the Kaluza-Klein metric ansatz. In 6-dimensional spacetime,  it is found that Einstein gravity with the cosmological constant  admits classical solutions with vanishing cosmological constant and the two extra dimensions are compacted. However, there also exist solutions with the non-vanishing cosmological constant. A vanishing cosmological constant is not preferred at the classical level. Finally, Canfora et al. studied the cosmological dynamics in LL gravity (with the Gauss-Bonnet term) in arbitrary dimensions \cite{can:2018}. So far, this is the review of the research work on LL gravity in higher dimensional cosmologies. 

This paper is also devoted to the study of higher dimensional cosmologies in the framework of LL gravity. Staring from the metric of four dimensional Robertson-Walker Universe plus an n-dimensional sphere with constant radius, we derive the equations of motion in LL gravity. Here we take into account three terms,  a large cosmological constant, the Einstein-Hilbert term and a higher order LL term except for the matter one in the action.  We find that the curvature of extra sphere brings us a huge but negative energy density if its radius is in the order of Planck length. Therefore, the large cosmological constant can be exactly balanced. In other words,  the huge vacuum energy bends the extra dimensional space into a Planck-length sized sphere while leaving the four dimensional physical spacetime flat.  On the other hand, the higher order LL term  results in a small cosmological constant which is consistent with the cosmic observations.  In order that the resulting  cosmological constant is small enough to explain the cosmic speed-up, the number of extra dimensions should be very large, such as $n\approx{69}$. Throughout this paper, we adopt the system of units in which $G=c=\hbar=1$ and the metric signature
$(-, +, +, +)$.

\section{The large and small cosmological constants: cancellation and generation}
Following Deruelle and Fari$\Tilde{\mathrm{n}}$a-Busto's research work \cite{der:1990},  we consider the line element  
\begin{equation}
ds^2=-dt^2+a\left(t\right)^2d\sigma_3^2+b^2d\sigma_{n}^2\;,
\end{equation}
where $d\sigma_3^2$  denotes our three dimensional Euclid space and $d\sigma_n^2$ the $n$-dimensional sphere for extra space. The scale factor $b$ of extra space is a constant because it has been compactified. The size of $b$ is commonly assumed to be several orders of magnitude of Planck length, $l_p$, 
\begin{eqnarray}
l_p=\sqrt{\frac{\hbar{G}}{c^3}}\approx 1.6\cdot 10^{-35}\mathrm{m}\;,
\end{eqnarray}
in order that it is consistent with gravitational and astronomical observations.  $a(t)$ is the scale factor of our four dimensional Universe. \blue{We note that the scale factor $b$ is in general also the function of cosmic time $t$. However, there are no problems with the assumption that it has been compactified as a constant which is frequently adopted in literature, for example, in Refs.~\cite{muller:1986,dem:1990,ark:1998}}. 
Our model has the following action 

\begin{eqnarray}
{S}=\int d^{4+n} x\sqrt{-g}\left[\frac{1}{16\pi}\left(-\frac{2\alpha_0}{l_p^2}+{\alpha_1}L_1+\frac{\alpha_N l_p^{2N-2}}{n_{2N-1}}L_{N}\right)+L_{m}\right]\;.\label{ll}
\end{eqnarray}
 Here $L_1$ and $L_N$ are defined by 
\begin{eqnarray}
L_1=R\;,\ \ \ L_N=2^{-N}\delta^{i_1i_2\cdot\cdot\cdot i_{2N}}_{j_1j_2\cdot\cdot\cdot j_{2N}}R^{j_{1}j_{2}}_{i_1i_2}\cdot\cdot\cdot R^{j_{2N-1}j_{2N}}_{i_{2N-1}i_{2N}}\;.
\end{eqnarray}
We see $L_1$ and $L_N$ represent the Ricci scalar and the $\mathrm{N}$ order LL invariant, respectively.  They have the dimension of $L^{-2}$ and $L^{-2N}$, respectively, because the Riemann curvature has the dimension of $L^{-2}$ ($L$ denotes the length). $\delta^{i_1i_2\cdot\cdot\cdot i_{2N}}_{j_1j_2\cdot\cdot\cdot j_{2N}}$ is the Kronecker symbol of order $2N$ and $R^{ij}_{ef}$ is the $(n+4)$-dimensional Riemann tensor. $n$ is the dimension of extra space and $\alpha_i$ are dimensionless constants. The term $-\alpha_0/l_p^2$ corresponds to the Einstein cosmological constant and it has the dimension of ${l_p^{-2}}$.  $L_1$ term is the contribution of Einstein-Hilbert term and the $L_N$ term comes from the higher order LL term. $L_m$  is the Lagrangian of matters. We note that different from the convention of Deruelle and Fari$\Tilde{\mathrm{n}}$a-Busto in reference \cite{der:1990},  a quotient coefficient of $n_{2N-1}$ arises in the front of $L_N$. Its subtlety lies in the fact that it helps create an effective  small cosmological constant.  General Relativity predicts $\alpha_1=1$.  $N$ is a positive integer and is constrained by the dimension of spacetime as follows  
\begin{eqnarray}\label{N1}
2<N\leq\frac{4+n}{2}\;,
\end{eqnarray}
when $n$ is an even number,  and 
\begin{eqnarray}\label{N2}
2<N\leq\frac{3+n}{2}\;,
\end{eqnarray}
when is odd. According to the convention of Deruelle and Fari$\Tilde{\mathrm{n}}$a-Busto in reference \cite{der:1990}, $n_1$ and $n_{2N-1}$ are defined by
\begin{eqnarray}
&&n_{1}\equiv{n\left(n-1\right)}\;,\ \ \ \  n_{2N-1}\equiv n\left(n-1\right)\left(n-2\right)\cdot\cdot\cdot\left[n-\left(2N-1\right)\right]\;.
\end{eqnarray}
Variation of action, Eq.~(\ref{ll}), with respect to the metric, one obtain the LL gravitational equations
\begin{eqnarray}
-\frac{\alpha_0}{l_p^2}g_{\mu\nu}+\alpha_1 G_{\mu\nu}+\frac{\alpha_Nl_p^{2N-2}}{n_{2N-1}}G^{(N)}_{\mu\nu}=8\pi T_{\mu\nu}\;,
\end{eqnarray}

with
\begin{eqnarray}
G^{(N)}_{\mu\nu}=-2^{-N-1}g_{\mu\rho}\delta^{\rho i_1i_2\cdot\cdot\cdot i_{2p}}_{\nu j_1j_2\cdot\cdot\cdot j_{2p}}R^{j_{1}j_{2}}_{i_1i_2}\cdot\cdot\cdot R^{j_{2p-1}j_{2p}}_{i_{2p-1}i_{2p}}\;.
\end{eqnarray}
$T_{\mu\nu}$ is the energy-momentum tensor in $n+4$ dimensional spacetime. For the cosmic fluid, it can be written as
\begin{eqnarray}
T_{\mu\nu}=\left[\rho\;,\ -pg_{ij}\;,\ -qg_{mn}\right]\;,
\end{eqnarray}
where $\rho$ and $p$ are the energy density and pressure in $3$-dimensional physical space and $q$ is the pressure in $n$-dimensional extra space. $g_{ij}$ and $g_{mn}$ denote the metric of physical space and extra space, respectively. 
   
Ref.~\cite{der:1990}  gives the equations of motion 

\begin{eqnarray}\label{fr}
-\frac{2\alpha_0}{l_p^2}+\alpha_1\left(6H^2+\frac{n_1}{b^2}\right)+\alpha_N\left[\left(\frac{l_p}{b}\right)^{2N}\cdot{\frac{1}{l_p^2}}+\frac{6Nn_{2N-3}}{n_{2N-1}}\cdot {H^2}\cdot\left(\frac{l_p}{b}\right)^{2N-2}\right]=16\pi\rho\;,
\end{eqnarray}

\begin{eqnarray}\label{acc1}
&&-\frac{2\alpha_0}{l_p^2}+\alpha_1\left(2H^2+\frac{n_1}{b^2}\right)+\alpha_N\left[\left(\frac{l_p}{b}\right)^{2N}\cdot{\frac{1}{l_p^2}}+\frac{2Nn_{2N-3}}{n_{2N-1}}\cdot{H^2}\cdot\left(\frac{l_p}{b}\right)^{2N-2}\right]\nonumber\\&&+4\frac{\ddot{a}}{a}\left[\alpha_1+\frac{N\alpha_Nn_{2N-3}}{n_{2N-1}}\cdot\left(\frac{l_p}{b}\right)^{2N-2}\right]=-16\pi{p}\;,
\end{eqnarray}

\begin{eqnarray}\label{acc2}
&&6\frac{\ddot{a}}{a}\left\{\alpha_1+N\alpha_N\left[\frac{\left(n-1\right)_{2N-2}}{n_{2N-1}}\cdot\left(\frac{l_p}{b}\right)^{2N-2}+\frac{2\left(N-1\right)\left(n-1\right)_{2N-4}{H^2l_p^2}}{n_{2N-1}}\cdot\left(\frac{l_p}{b}\right)^{2N-4}\right]\right\}\nonumber\\&&-\frac{2\alpha_0}{l_p^2}+\alpha_N\left[\frac{\left(n-1\right)_{2N}}{n_{2N-1}}\cdot\left(\frac{l_p}{b}\right)^{2N}\cdot{\frac{1}{l_p^2}}+\frac{6N\left(n-1\right)_{2N-2}}{n_{2N-1}}\cdot{H^{2N-2}}\cdot\left(\frac{l_p}{b}\right)^2\right]\nonumber\\&&\nonumber\\&&+\alpha_1\left[6H^2+\frac{\left(n-1\right)_2}{b^2}\right]=-16\pi{q}\;,
\end{eqnarray}
where the dot denotes the derivative with respect to cosmic time, $t$. $H$ is the Hubble parameter which has the dimension of $L^{-1}$. Now we have four variables, $a, \rho, p$ and $q$, but three in-dependant equations. The four in-dependent equation is needed. It is the equation of state for matters,
\begin{eqnarray}
p=F\left(\rho\;, q\right)\;.
\end{eqnarray} 
If the scale factor $b$ of extra space is also time-dependant, the energy conservation equation 
\begin{eqnarray}
T^{\mu\nu}_{;\nu}=0\;,
\end{eqnarray}
turns out to be
\begin{eqnarray}
\frac{d\rho}{dt}+3\frac{\dot{a}}{a}\left(\rho+p\right)+n\frac{\dot{b}}{b}\left(\rho+q\right)=0\;.
\end{eqnarray}
Now the scale factor $b$ is a constant, we have 
\begin{eqnarray}\label{ece}
\frac{d\rho}{dt}+3\frac{\dot{a}}{a}\left(\rho+p\right)=0\;.
\end{eqnarray}
In this case, Eq.~(\ref{acc1}) can be replaced with Eq.~(\ref{ece}). As a result, Eq~(\ref{fr}),  Eq.~(\ref{ece}) plus the equation for equation of state 
\begin{eqnarray}
p=F\left(\rho\right)\;.
\end{eqnarray}
constitute the closed system of equations of motion.  We do not need to consider Eq.~(\ref{acc2}) any more.

Then Friedmann equation Eq.~(\ref{fr}) can be rewritten as  
\begin{eqnarray}\label{fra}
\alpha_1 3H^2=8\pi\rho+\frac{\alpha_0}{l_p^2}-\alpha_1\frac{n_1}{2b^2}-\frac{1}{2}\alpha_N\left[\left(\frac{l_p}{b}\right)^{2N}\cdot{\frac{1}{l_p^2}}+\frac{6Nn_{2N-3}H^2}{n_{2N-1}}\cdot\left(\frac{l_p}{b}\right)^{2N-2}\right]\;.
\end{eqnarray}
In order that it restores to the Friedmann equation in General Relativity, we must require
\begin{eqnarray}
\alpha_1=1\;.
\end{eqnarray}
\blue{Now let's make an analysis of the terms on the right  hand side of the equation. The first term is cosmic energy density (or mass density) contributed mainly by dark matter and relativistic  matter.  For the present-day Universe, it is roughly} \blue{\begin{eqnarray}
\rho\longrightarrow\frac{\rho}{\rho_p}\cdot\frac{1}{l_p^2}=10^{-124}\cdot\frac{1}{l_p^2}\;,
\end{eqnarray}} 
\blue{in the unit of inverse of length squared. $\rho_p$ is called the Planck mass density. If the dimensionless constant $\alpha_0$ is in the order of unit one $\alpha_0\approx{1}$, we find the ratio of first term to second term is about $10^{-124}$, an extremely small number. It is also the case for the third term if we assume $\alpha_1$ is in the order of unit one, $n_1$ not a large number and $b$ a few order of Planck length. The second term and the third term have different signs. This inspires us the second term contributes the huge vacuum energy and the third term is used to counteract it. As for the fourth and fifth term in square bracket, they can be in the order of $10^{-124}$ provided that $N$ is sufficiently large. This allows us to interpret the two terms as dark energy. We shall explain it in the following manner.}

If the cosmological constant is understood as the mass density of vacuum, the estimation of quantum field theory gives the following magnitude for dimensionless constant $\alpha_0$
\begin{eqnarray}
\alpha_0\left(\mathrm{theory}\right)\approx{1}\;.
\end{eqnarray}
When \blue{the second term} is transformed to the unit of mass density, we find
\blue{\begin{eqnarray}
\frac{\alpha_0\left(\mathrm{theory}\right)}{l_p^2}=\frac{1}{l_p^2}\rightarrow\frac{l_p^2}{l_p^2}\cdot\rho_{p}={\frac{c^5}{\hbar{G}^2}}\approx 5.2\cdot{10^{96}}\mathrm{kg}\cdot\rm{\dot{m^{-3}}}\;.
\end{eqnarray}}
Apart from numerical factors this density is the sum of zero-point energies of oscillation from the electromagnetic field.  However, assuming that dark energy in the present-day Universe  is contributed by the cosmological constant, we are led to the following value   
\begin{eqnarray}
\alpha_0\left(\mathrm{observations}\right)\approx 10^{-124}\;,
\end{eqnarray}
by various observations. In the unit of mass density, it is 

\blue{\begin{eqnarray}
\frac{\alpha_0\left(\mathrm{observations}\right)}{l_p^2}=\frac{10^{-124}}{l_p^2}\rightarrow\rho_{X}=10^{-124}\cdot\frac{l_p^2}{l_p^2}\cdot\rho_p=10^{-124}{\frac{c^5}{\hbar{G}^2}}\approx 5.2\cdot{10^{-28}}\mathrm{kg}\cdot\rm{\dot{m^{-3}}}\;.
\end{eqnarray}}
Thus we see that the theoretical result $\alpha_0\approx 1$ is the order of $10^{124}$ of the observational value $\alpha_0\approx 10^{-124}$. This is the famous cosmological constant problem. We note that the term $n_1$ has the dimension of $L^{-2}$ . If the scale of extra space is in the order of Planck length, the $n_1$ term  would contribute a huge but negative energy density, i.e.
\blue{\begin{eqnarray}
-\alpha_1\frac{n_1}{2b^2}=-\frac{n_1}{2l_p^2}\rightarrow-\frac{n_1}{2}\cdot\frac{l_p^2}{l_p^2}\cdot\rho_p=-\frac{n_1}{2}\rho_p\;.
\end{eqnarray}}
This tells us the huge vacuum energy can be interacted by the curvature of extra sphere. If we let 
\blue{\begin{eqnarray}
\frac{\alpha_0\left(\mathrm{theory}\right)}{l_p^2}-\frac{n_1}{2b^2}\rightarrow{10^{-124}}\cdot\frac{1}{l_p^2}\;,
\end{eqnarray}}
this means that we are required to adjust the counter term of $\frac{n_1}{2b^2}$ to at least one part in $10^{124}$.  This amounts to the accuracy that we require a weight difference of $10^{-114}\mathrm{kg}$ between two mountains. Thus an extremely  fine-tuning problem appears. 
Therefore, we demand 
\blue{\begin{eqnarray}
\frac{\alpha_0\left(\mathrm{theory}\right)}{l_p^2}-\frac{n_1}{2b^2}=0\;.
\end{eqnarray}}
Now the huge positive cosmological constant is exactly  neutralized by negative energy of curvature of extra space.  We then obtain
\blue{\begin{eqnarray}\label{B}
b=\sqrt{\frac{n_1}{2\alpha_0\left(\mathrm{theory}\right)}}l_p\;.
\end{eqnarray}}
 This reveals that in order to counteract the huge  vacuum energy by the curvature of extra space, the size of extra space could be several orders of Planck length. The scale of extra space $b$ is proportional to $\sqrt{n_1}$.  This means the radius $b$ increases with the increasing of dimension $n$. If we assume the scale of extra space is smaller than that of quark particles, i.e.
 \begin{eqnarray}
b<10^{-18}\mathrm{m}\;,
\end{eqnarray}
we find 
\begin{eqnarray}
n<10^{17}\;.
\end{eqnarray} 
Namely, the dimension of extra space can not be arbitrarily large.   
 Substituting the scale of extra space, Eq.~(\ref{B})  into Eq.~(\ref{fra}), we obtain the density of dark energy 

\begin{eqnarray}\label{DE}
\rho_X=-\frac{1}{16\pi}\alpha_N\left[\left(\frac{2\alpha_0}{n_1}\right)^{N}\cdot{\rho_p}+\frac{6Nn_{2N-3}H^2}{n_{2N-1}}\cdot\left(\frac{2\alpha_0}{n_1}\right)^{N-1}\right]\;.
\end{eqnarray}
\blue{Since $\alpha_0\left(\mathrm{theory}\right)\approx{1}$}, we conjecture the dimensionless constant $\alpha_N$ is also in the order of unit one, i.e. $\alpha_N\approx{-1}$ and we let
\begin{eqnarray}
\frac{1}{16\pi}\left(\frac{2}{n_1}\right)^{N}=10^{-124}\;.
\end{eqnarray}
As indicated earlier, the maximum of $N$ is $\frac{n+4}{2}$ or $\frac{n+3}{2}$. Considering this situation,  we have  
\begin{eqnarray}
\frac{1}{16\pi}\left[\frac{2}{n\left(n-1\right)}\right]^{\frac{n+4}{2}}=10^{-124}\;,\ \ \ \ or \ \ \ \frac{1}{16\pi}\left[\frac{2}{n\left(n-1\right)}\right]^{\frac{n+3}{2}}=10^{-124}\;.
\end{eqnarray}
Therefore, we conclude the number of extra spatial dimensions is
\begin{eqnarray}\label{numextra}
n\approx{69}\;,
\end{eqnarray}
in order that the density of dark energy is in the order 
\begin{eqnarray}
\rho_X\approx 10^{-124}\rho_p\;.
\end{eqnarray}
\blue{For convenience, the symbol $\alpha_0\left(\mathrm{theory}\right)$ would be be abbreviated to $\alpha_0$ in the following}. One can check that in the present-day Universe,  the second term is very much smaller than the first one by making an examination on their ratio 
\begin{eqnarray}\label{eq2}
\mathcal{R}\equiv\frac{\frac{6Nn_{2N-3}H_0^2}{n_{2N-1}}\cdot\left(\frac{2\alpha_0}{n_1}\right)^{N-1}}{\left(\frac{2\alpha_0}{n_1}\right)^{N}\cdot{\rho_p}}=\frac{8{\pi} Nn\left(n-1\right)}{\alpha_0\left(n-2N+1\right)\left(n-2N+2\right)}\cdot{\frac{\rho_0}{\rho_p}}\;,
\end{eqnarray}
where 
\begin{eqnarray}
\rho_0=\frac{8{\pi}H_0^2}{3}\;,
\end{eqnarray}
is the present-day cosmic mass density. $H_0$ is the Hubble parameter today. Substituting  \begin{eqnarray}
n=69\;,\ \ \ N=36\ (\mathrm{or\ \ } 36.5)\;,\ \ \  \alpha_0=1\;,\ \  \rho_0={10^{-28}}\mathrm{kg}\cdot\rm{\dot{m^{-3}}}\;,
\end{eqnarray}
into above equation, we obtain   
\begin{eqnarray}
\mathcal{R}=10^{-119}\ll{1}\;,
\end{eqnarray}
Therefore, the second term can be safely neglected. Eventually, we obtain the expression for a small cosmological constant
\begin{eqnarray}\label{DE}
\rho_X=-\frac{1}{16\pi}\alpha_N\left(\frac{2\alpha_0}{n_1}\right)^{N}\rho_p\;.
\end{eqnarray}
It is apparent $\alpha_n$ is required to be negative in order for the energy density to be positive. We expect $\alpha_N$ is the order of unit one.  We note that, in order to get an observations allowed dark energy density, we may not select the maximum of $N$ if the dimension of extra space is sufficiently large, for  example, the pairs of $n=10^{7}$ with $N=9$,  $n=10^{9}$ with $N=7$, $n=10^{12}$ with $N=5$, and so on.

\section{conclusion and discussion}
To sum up, we have proposed a mechanism that  cancels out the huge  cosmological constant and a small cosmological constant is produced. Our strategy is to bend the extra space into an n-dimensional sphere with a constant radius of order of Planck length.  To achieve this,  a huge positive energy density is needed. This is precisely the role of vacuum energy.  To put it in another way, the n-dimensional sphere contains huge but negative energy density and it exactly neutralize the huge and positive vacuum energy.  In the mean time an observational allowed small cosmological constant is produced.   In this instance, a higher order LL term is needed.  If  the number of extra dimensions is sufficiently large, such as $n\approx{69}$, the scale of extra space would be approximately  $6l_p$. \blue{Up to this point our limit on the number of extra dimensions is based on the LL gravity. In contrast, Lin et al \cite{lin:2024}, in the framework of Einstein gravity, provide another novel way to limit the number of extra dimensions in the braneworld model by using the observations of gravitational waves}.     

Why can the huge vacuum energy density be cancelled out by extra space? From the perspective of dimensional analysis,  we see the vacuum energy density is proportional to $l_p^{-2}$. It is subtly that the extra sphere contains a minus energy tensity which is similarly proportional to $b^{-2}$.  If the scale of extra space, $b$ , is the order of Planck length,   the two energy densities can be precisely offset.  On the other hand, why can a small cosmological constant be produced by a large number  of extra dimensions? Also  from the perspective of dimensional analysis, we see the density of dark energy is proportional to $(\frac{l_p}{b})^{2N}$ or ${b}^{-2N}$ from Eq.~({\ref{fra}}). This is exactly the dimension of $\mathrm{N-th}$ order LL invariant. Since $b>l_p$, we see the larger magnitude the index $\mathrm{N}$, the smaller magnitude the energy density.   Only with a sufficiently large number of extra dimensions can we obtain a sufficiently large N.  So a large number of extra dimensions are required in order to get a small cosmological constant.
     
\section{\bf Acknowledgments}

The work is supported by the National Key R$\&$D Program of China grants No. 2022YFF0503404 and No. 2022SKA0110100.

\vspace*{0.6cm}
%\section*{References}
\newcommand\arctanh[3]{~arctanh.{\bf ~#1}, #2~ (#3)}
\newcommand\ARNPS[3]{~Ann. Rev. Nucl. Part. Sci.{\bf ~#1}, #2~ (#3)}
\newcommand\AL[3]{~Astron. Lett.{\bf ~#1}, #2~ (#3)}
\newcommand\AP[3]{~Astropart. Phys.{\bf ~#1}, #2~ (#3)}
\newcommand\AJ[3]{~Astron. J.{\bf ~#1}, #2~(#3)}
\newcommand\GC[3]{~Grav. Cosmol.{\bf ~#1}, #2~(#3)}
\newcommand\APJ[3]{~Astrophys. J.{\bf ~#1}, #2~ (#3)}
\newcommand\APJL[3]{~Astrophys. J. Lett. {\bf ~#1}, L#2~(#3)}
\newcommand\APJS[3]{~Astrophys. J. Suppl. Ser.{\bf ~#1}, #2~(#3)}
\newcommand\JHEP[3]{~JHEP.{\bf ~#1}, #2~(#3)}
\newcommand\JMP[3]{~J. Math. Phys. {\bf ~#1}, #2~(#3)}
\newcommand\JCAP[3]{~JCAP {\bf ~#1}, #2~ (#3)}
\newcommand\LRR[3]{~Living Rev. Relativity. {\bf ~#1}, #2~ (#3)}
\newcommand\MNRAS[3]{~Mon. Not. R. Astron. Soc.{\bf ~#1}, #2~(#3)}
\newcommand\MNRASL[3]{~Mon. Not. R. Astron. Soc.{\bf ~#1}, L#2~(#3)}
\newcommand\NPB[3]{~Nucl. Phys. B{\bf ~#1}, #2~(#3)}
\newcommand\CMP[3]{~Comm. Math. Phys.{\bf ~#1}, #2~(#3)}
\newcommand\CQG[3]{~Class. Quantum Grav.{\bf ~#1}, #2~(#3)}
\newcommand\PLB[3]{~Phys. Lett. B{\bf ~#1}, #2~(#3)}
\newcommand\PRL[3]{~Phys. Rev. Lett.{\bf ~#1}, #2~(#3)}
\newcommand\PR[3]{~Phys. Rep.{\bf ~#1}, #2~(#3)}
\newcommand\PRd[3]{~Phys. Rev.{\bf ~#1}, #2~(#3)}
\newcommand\PRD[3]{~Phys. Rev. D{\bf ~#1}, #2~(#3)}
\newcommand\RMP[3]{~Rev. Mod. Phys.{\bf ~#1}, #2~(#3)}
\newcommand\SJNP[3]{~Sov. J. Nucl. Phys.{\bf ~#1}, #2~(#3)}
\newcommand\ZPC[3]{~Z. Phys. C{\bf ~#1}, #2~(#3)}
\newcommand\IJGMP[3]{~Int. J. Geom. Meth. Mod. Phys.{\bf ~#1}, #2~(#3)}
\newcommand\IJTP[3]{~Int. J. Theo. Phys.{\bf ~#1}, #2~(#3)}
\newcommand\IJMPD[3]{~Int. J. Mod. Phys. D{\bf ~#1}, #2~(#3)}
\newcommand\IJMPA[3]{~Int. J. Mod. Phys. A{\bf ~#1}, #2~(#3)}
\newcommand\GRG[3]{~Gen. Rel. Grav.{\bf ~#1}, #2~(#3)}
\newcommand\EPJC[3]{~Eur. Phys. J. C{\bf ~#1}, #2~(#3)}
\newcommand\PRSLA[3]{~Proc. Roy. Soc. Lond. A {\bf ~#1}, #2~(#3)}
\newcommand\AHEP[3]{~Adv. High Energy Phys.{\bf ~#1}, #2~(#3)}
\newcommand\Pramana[3]{~Pramana.{\bf ~#1}, #2~(#3)}
\newcommand\PTEP[3]{~PTEP.{\bf ~#1}, #2~(#3)}
\newcommand\PTP[3]{~Prog. Theor. Phys{\bf ~#1}, #2~(#3)}
\newcommand\APPS[3]{~Acta Phys. Polon. Supp.{\bf ~#1}, #2~(#3)}
\newcommand\ANP[3]{~Annals Phys.{\bf ~#1}, #2~(#3)}
\newcommand\RPP[3]{~Rept. Prog. Phys. {\bf ~#1}, #2~(#3)}
\newcommand\ZP[3]{~Z. Phys. {\bf ~#1}, #2~(#3)}
\newcommand\NCBS[3]{~Nuovo Cimento B Serie {\bf ~#1}, #2~(#3)}
\newcommand\AAP[3]{~Astron. Astrophys.{\bf ~#1}, #2~(#3)}
\newcommand\MPLA[3]{~Mod. Phys. Lett. A.{\bf ~#1}, #2~(#3)}
\newcommand\NT[3]{~Nature.{\bf ~#1}, #2~(#3)}
\newcommand\PT[3]{~Phys. Today. {\bf ~#1}, #2~ (#3)}
\newcommand\APPB[3]{~Acta Phys. Polon. B{\bf ~#1}, #2~(#3)}
\newcommand\NP[3]{~Nucl. Phys. {\bf ~#1}, #2~ (#3)}
\newcommand\JETP[3]{~JETP Lett. {\bf ~#1}, #2~(#3)}
\newcommand\PDU[3]{~Phys. Dark. Univ. {\bf ~#1}, #2~(#3)}
\newcommand\UNI[3]{~Universe. {\bf ~#1}, #2~(#3)}
\newcommand\OAJ[3]{~The Open Astronomy Journal. {\bf ~#1}, #2~(#3)}
\newcommand\LNP[3]{~Lect. Notes Phys.  {\bf ~#1}, #2~(#3)}
\newcommand\ANM[3]{~Annals. Math.{\bf ~#1}, #2~(#3)}
\newcommand\AM[3]{~Aequationes Math. {\bf ~#1}, #2~(#3)}

\newcommand\SPAW[3]{~Sitzungober. Preuss. Akad. Wiss. Berlin. {\bf ~#1}, #2~(#3)}

\end{document}